\documentclass[final,5p,times]{elsarticle}

\usepackage{xcolor}

\usepackage{amssymb}
\usepackage{graphics}
\usepackage{xcolor}
\usepackage{sidecap}
\usepackage{microtype}
\usepackage{graphicx}
\usepackage{dcolumn}
\usepackage{bm}
\usepackage[separate-uncertainty=true]{siunitx}
\usepackage{hyperref}
\usepackage{float}
\usepackage{tikz}
\usepackage{caption}
\usepackage{subcaption}
\usepackage{comment}

\journal{Physics Letters B}

\begin{document}
\begin{frontmatter}

\title{Photoproduction of $K^+\Lambda(1405) \rightarrow K^+\pi^0\Sigma^0$ extending to forward angles and low momentum transfer }
\author[1]{G.~Scheluchin}
\author[1]{T.C.~Jude\corref{cor1}}
\ead{jude@physik.uni-bonn.de}
\author[1]{S.~Alef}
\author[1]{P.~Bauer}
\author[2,3]{D.~Bayadilov}
\author[2]{R.~Beck}
\author[4]{A.~Braghieri}
\author[5]{P.L.~Cole}
\author[1]{D.~Elsner}
\author[6]{R.~Di Salvo}
\author[6,7]{A.~Fantini}
\author[1]{O.~Freyermuth}
\author[1]{F.~Frommberger}
\author[8,9]{F.~Ghio}
\author[3]{A.~Gridnev}
\author[1]{D.~Hammann\fnref{fn1} }%
\author[1]{J.~Hannappel \fnref{fn2} }
\author[1]{K.~Kohl}
\author[3]{N.~Kozlenko}
\author[10]{A.~Lapik}
\author[11]{P.~Levi Sandri}
\author[10]{V.~Lisin}
\author[12,13]{G.~Mandaglio}
\author[6,7]{R.~Messi}
\author[11]{D.~Moricciani  \fnref{fn3}}
\author[10]{A.~Mushkarenkov}
\author[10]{V.~Nedorezov}
\author[3]{D.~Novinskiy}
\author[4]{P.~Pedroni}
\author[10]{A.~Polonskiy}
\author[1]{B.-E.~Reitz \fnref{fn1} }%
\author[6,14]{M.~Romaniuk}
\author[1]{H.~Schmieden}
\author[3]{V.~Sumachev  \fnref{fn3} }
\author[3]{V.~Tarakanov}

\cortext[cor1]{Corresponding author}
\fntext[fn1]{No longer employed in academia}
\fntext[fn2]{Currently, DESY Research Centre, Hamburg, Germany}
\fntext[fn3]{Deceased}

\address[1]{Rheinische Friedrich-Wilhelms-Universit\"at Bonn, Physikalisches Institut, Nu\ss allee 12, 53115 Bonn, Germany}
\address[2]{Rheinische Friedrich-Wilhelms-Universit\"at Bonn, Helmholtz-Institut f\"ur Strahlen- und Kernphysik, Nu\ss allee 14-16, 53115 Bonn, Germany}
\address[3]{Petersburg Nuclear Physics Institute NRC  ``Kurchatov Institute", Gatchina, Leningrad District, 188300, Russia}
\address[4]{INFN sezione di Pavia, Via Agostino Bassi, 6 - 27100 Pavia, Italy}
\address[5]{Lamar University, Department of Physics, Beaumont, Texas, 77710, USA}
\address[6]{INFN Roma ``Tor Vergata", Via della Ricerca Scientifica 1, 00133, Rome, Italy}
\address[7]{Universit\`a di Roma ``Tor Vergata'', Dipartimento di Fisica, Via della Ricerca Scientifica 1, 00133, Rome, Italy}
\address[8]{INFN sezione di Roma La Sapienza, P.le Aldo Moro 2, 00185, Rome, Italy}
\address[9]{Istituto Superiore di Sanit\`a, Viale Regina Elena 299, 00161, Rome, Italy}
\address[10]{Russian Academy of Sciences Institute for Nuclear Research, Prospekt 60-letiya Oktyabrya 7a, 117312, Moscow, Russia}
\address[11]{INFN - Laboratori Nazionali di Frascati, Via E. Fermi 54, 00044, Frascati, Italy}
\address[12]{INFN sezione Catania, 95129, Catania, Italy}
\address[13]{Universit\`a degli Studi di Messina, Dipartimento MIFT,  Via F. S. D'Alcontres 31, 98166, Messina, Italy}
\address[14]{Institute for Nuclear Research of NASU, 03028, Kyiv, Ukraine}


\date{\today}

\begin{abstract}
$K^+\Lambda(1405)$ photoproduction has been studied at the BGOOD experiment via the all neutral decay, $\Lambda(1405)\rightarrow\Sigma^0\pi^0$. 
The unique BGOOD experimental setup allows both the cross section and $\Lambda(1405)$ invariant mass distribution (line shape) to be measured over a broad $K^+$ polar angle range, extending to extreme forward $K^+$ angles unattainable at previous experiments.

Evidence is provided for the role of a triangle singularity driven by the $N^*(2030)$ resonance, which appears to contribute significantly  to $K^+\Lambda(1405)$ photoproduction.
This is observed in both the angular distributions and the integrated cross section which was determined with unprecedented energy resolution.  The measured line shape is also in agreement with the previous results of CLAS and ANKE, and is consistent with two poles derived in $\chi$PT based models. 

\end{abstract}


\end{frontmatter}

\section{Introduction}

The $\Lambda(1405)$ resonance is a strangeness $S=-1$ baryon with spin-parity $J^P=1^-$ and isospin $I=0$, situated directly at the $\bar K N$ threshold and decaying into the three $\pi \Sigma$ channels through the strong interaction. 
Since its discovery in bubble chamber studies of the reaction $K^- p \rightarrow \pi \pi \pi \Sigma$~\cite{bib:Alston}, it has attracted a lot of experimental activity in $K^- p$ scattering~\cite{bib:Hemingway,bib:pdg}, in $pp$ scattering~\cite{bib:anke_lineshape,bib:AgakishievHADES}, in the study of kaonic hydrogen~\cite{bib:Bazzi1,bib:Bazzi2}, and in $K^+ \Lambda(1405)$ photoproduction~\cite{bib:NiiyamaLEPS,bib:MoriyaSpin,bib:clas_crosssection}. 

Interestingly, when discovered prior to the advent of the quark model, the $\Lambda(1405)$ was predicted to be a $\bar K N$ molecular-type bound state~\cite{bib:dalitz}, mainly because it is located directly below the $\bar K N$ threshold. 
The invariant mass distribution in the $(\pi \Sigma)^0$ decay, often called the \textit{line shape}, appears distorted from a usual Breit-Wigner resonance shape. The $\Lambda(1405)$ has recently achieved the status of the archetypal molecular hadronic state in the $uds$ sector~\cite{bib:2017jvc}, which is mainly due to new developments in Chiral Perturbation Theory ($\chi$PT) and Quantum Chromodynamics on the Lattice (LQCD). 
The former describe the $\Lambda(1405)$ through meson-baryon dynamics at the $\bar K$-nucleon threshold~\cite{bib:OsetRamos,bib:oller,bib:twopeak,bib:twopeakwidth,bib:mai}. 
The latter determine $\bar K$-nucleon as the dominating Fock component of the wave function when quark masses approach realistically small values~\cite{bib:latticeQCD,bib:henhart}. 
Such a molecular structure is difficult to reconcile with constituent quark models, which include the $\Lambda(1405)$ as the $uds$ flavour singlet 3-quark state~\cite{bib:Isgur}. 
Mass degeneracy with the anticipated spin-parity partner $\Lambda(1520)$ is not observed, and the non-strange $J^P=1^-$ partner $N(1535)$ is 130\,MeV heavier, despite lacking the valence strange quark. 
Within baryon $\chi$PT, a two pole structure is predicted in the $I=0$ channel~\cite{bib:twopeak}.
In the $\Lambda(1405)$ decay the three $(\pi \Sigma)^0$ charge combinations are expected to couple differently to the two poles which 
was experimentally confirmed in $K^+\Lambda(1405)$ photoproduction~\cite{bib:clas_lineshape}. 

\begin{figure} [htbp!]
	
	\vspace*{0cm}
	\resizebox{1\columnwidth}{!}{%
		
		\includegraphics{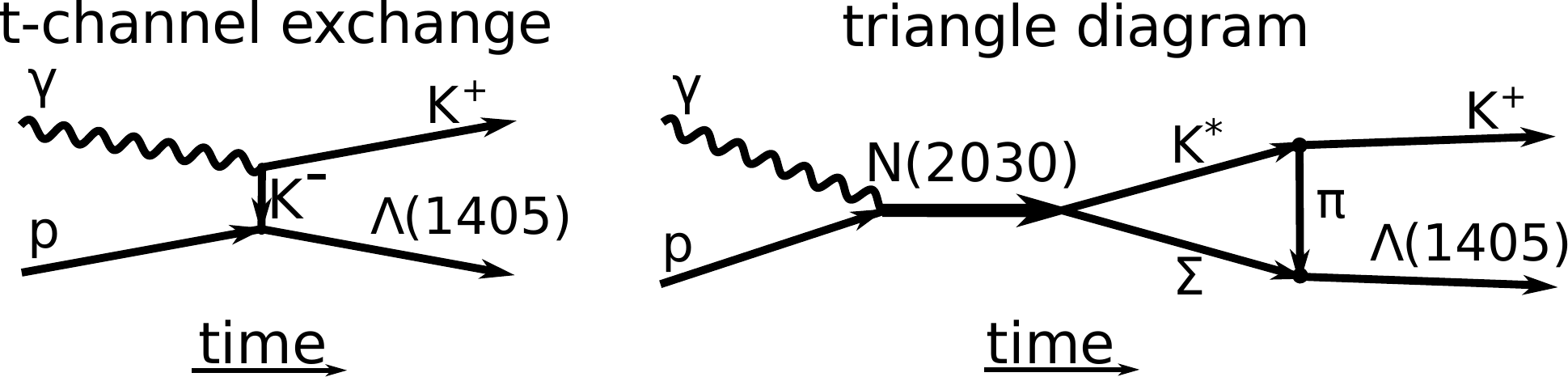} 
		
	}
	\caption{Possible processes of $\Lambda(1405)$ photoproduction.}
	
	\label{fig:tprocess}
\end{figure}

The reaction  $\gamma p \rightarrow K^+ \Lambda(1405)$ is assumed to mainly proceed via the kaon $t$-channel exchange depicted in Fig.~\ref{fig:tprocess} (left). 
Such processes are generally expected to be dominated by small momentum transfers, $t$, in particular if the $\Lambda(1405)$ has a relatively loosely bound molecular structure.
In this case, production would be prohibited if the momentum transfer is significantly larger than the internal momentum due to the Fermi motion of the hadron constituents within the molecule. 
Since small $t$ is associated with forward going $K^+$, corresponding forward angle acceptance is experimentally mandatory to study such possible processes. 

Ref.~\cite{bib:wang} suggests an additional mechanism, where production is driven via the triangle diagram shown in Fig.~\ref{fig:tprocess} (right). 
This is of particular interest not only regarding the photoproduction mechanism itself, but also in view of the existence and internal structure of the $s$-channel $N^*(2030)$ which drives the $\Lambda(1405)$ final state. 
The Coleman-Norton theorem~\cite{bib:Coleman} explains how the $N^*(2030)$ must have a mass close to the total free mass of the $K^* \Sigma$ and have a strong coupling to $K^* \Sigma$ which provide the internal legs of the triangle. 
If such a process contributes, it is likely that the $N^*(2030)$ is itself a molecular-type $K^* \Sigma$ state. 
And indeed, it is exactly this state which was held responsible for a cusp-like structure observed in $K^0 \Sigma^+$ photoproduction at the $K^* \Sigma$ threshold~\cite{bib:ewald} in the same $\chi$PT model of vector meson-baryon interactions~\cite{bib:OsetRamos2} which provided the only prediction of the pentaquarks~\cite{bib:juunwu} observed by LHCb~\cite{bib:LHCbpenta}. 
If the triangle mechanism provides a significant contribution to $\Lambda(1405)$ photoproduction, then, in contrast to a pure $t$-channel process, the differential cross section is suggested to flatten and drop at the most forward angles. 
An enhanced total cross section is expected in the region of the driving $N^*(2030)$ structure, which quickly drops once the $K^* \Sigma$ threshold is exceeded, whereby the triangle breaks up and the mechanism is lost for $\Lambda(1405)$ production. 

The BGOOD experiment~\cite{bib:teckpaper} at the ELSA electron accelerator~\cite{bib:elsa} at the University of Bonn is ideally suited to investigate these issues. The very forward $K^+$ acceptance and the almost hermetic acceptance to identify $\Lambda(1405) \rightarrow \pi \Sigma$ allows access to hitherto unexplored kinematic regions.
One of the challenges in extracting a clean selection of  $K^+\Lambda(1405)$ events is the near mass degeneracy between the $\Lambda(1405)$ and the $\Sigma(1385)$, which also decays to $\pi\Sigma$. 
Fortunately, $\Lambda(1405)\rightarrow \pi^0\Sigma^0$ is the exception since $\Sigma(1385)\rightarrow\pi^0\Sigma^0$ is prohibited by isospin conservation. 

This letter presents the differential and integrated photoproduction cross sections and the $\Lambda(1405)$ lineshape via the decay $\Lambda(1405) \rightarrow\pi^0\Sigma^0$.

%


\section{Experimental setup and analysis procedure}\label{sec:setup}

BGOOD~\cite{bib:teckpaper} is comprised of two main parts.  The central region is covered by the \textit{BGO Rugby Ball} calorimeter, which is ideal for neutral meson detection and complemented by inner  sub-detectors for charged particle identification.  The \textit{Forward Spectrometer} covers polar angles 1-12$^\circ$ for charged particle identification and momentum reconstruction.  The small intermediate region is covered by \textit{SciRi}, which consists of concentric rings of plastic scintillators for charged particle detection. 

The presented data  was taken over 49 days using a 6\,cm long liquid hydrogen target and an ELSA electron beam energy of 3.2\,GeV.
The electron beam was incident upon a \SI{560}{\mu m} thick diamond radiator\footnote{A diamond radiator was used to produce coherent, linearly polarised photon beam with a maximum polarisation at a beam energy of 1.4\,GeV, however the polarisation was not required for the presented analysis.} to produce an energy tagged bremsstrahlung photon beam which was subsequently collimated.  The photon beam energy, $E_\gamma$, was determined per event by momentum analysing the post bremsstrahlung electrons in the \textit{Photon Tagger} over the range of \SI{10}{\%} to \SI{90}{\%} of the accelerator energy.  The integrated photon flux from 1550 to 2900\,MeV was $8.5 \times 10^{12}$. 


The reaction channel was identified via the sequential decay $\gamma p \rightarrow K^+ \Lambda(1405) \rightarrow K^+ \pi^0 \Sigma^0 \rightarrow K^+ (\gamma\gamma)(\gamma\Lambda) \rightarrow K^+ (\gamma\gamma)(\gamma \pi^- p)$, therefore candidate events with three neutral and three charged particles were selected.
As it is difficult to distinguish between charged particles in the BGO Rugby Ball and SciRi, all combinations of each candidate event was determined, where the $K^{+}$, $\pi^-$ and proton were interchangeable between the measured tracks.  The exception to this was the Forward Spectrometer, where particles were identified via their mass reconstruction and combinatorial background was therefore suppressed.  For forward going $K^+$, this corresponds to $\cos(\theta)>0.86$, where $\theta$ is the $K^+$ polar angle in the centre-of-mass frame.

A kinematic fit was applied to all combinations, with the constraints of four-momentum conservation and the $\pi^0$ and $\Lambda$ invariant masses.  
As $\pi^-$ and protons can not be accurately distinguished in the BGO Rugby Ball and SciRi, the combination with the best confidence level from the kinematic fit was used and events with a confidence level lower than 0.2 were excluded from  further analysis to improve the signal to background ratio.

\begin{figure}[h]
	\centering
	\includegraphics[width=\columnwidth]{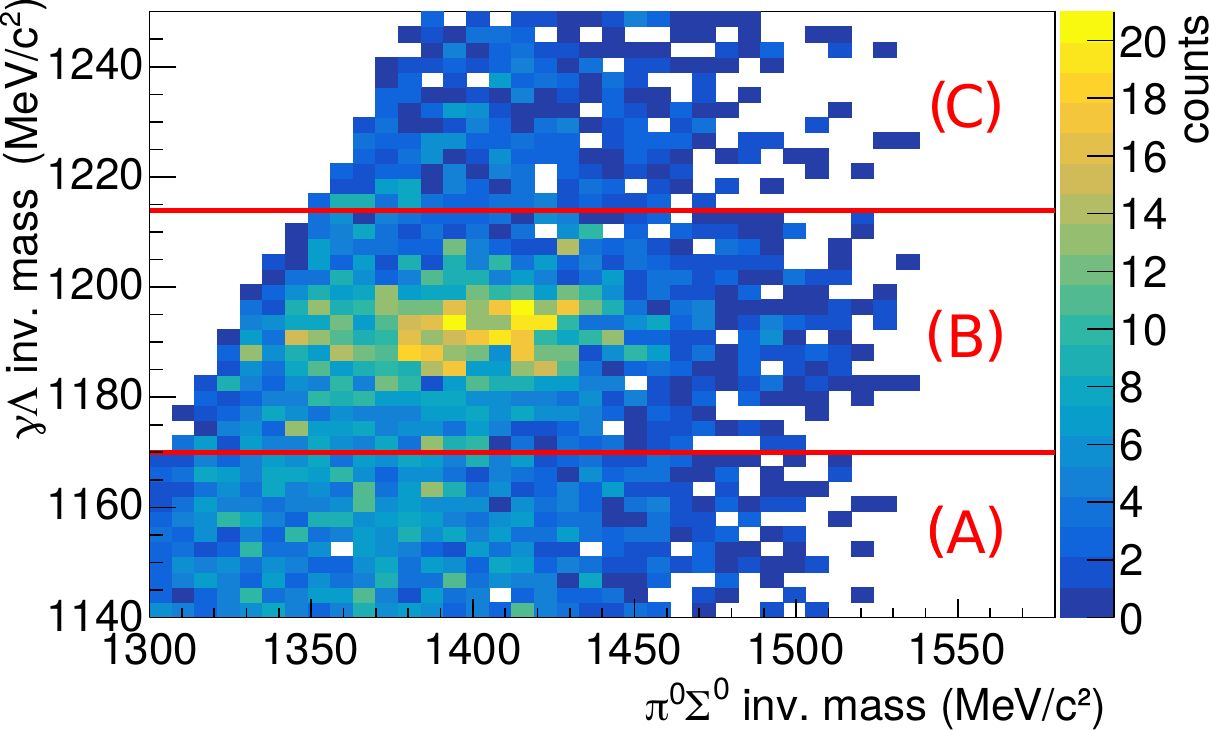} 
	\caption{$\gamma\Lambda$ versus $\pi^0\Sigma^0$ invariant mass distribution for $E_\gamma=$ 1550 to 1750\,MeV and $\cos(\theta)=$\SIrange{-1}{1}{}.  The sections labelled (A), (B) and (C) correspond to regions with different ratios of signal to background, described in the text.}\label{fig:RooFits_lineshape}
	\label{fig:RooFits_lineshape-1}
\end{figure}

Fig.~\ref{fig:RooFits_lineshape-1} shows the two dimensional plot of the  $\pi^{0}\Sigma^{0}$ and $\gamma \Lambda$ invariant mass.
The peak at approximately 1400\,MeV/c$^2$ and 1190\,MeV/c$^2$ corresponds to the $\Lambda(1405)$ and $\Sigma^0$ respectively.

To extract the line shape, simulated data was used to determine the background from reactions which do not decay to $K^+ \gamma \pi^- p \gamma \gamma$.
This  includes $\eta\pi^0p$, $K^0\Sigma^+$, $K^+\Sigma^0$, $\pi^0\pi^+\pi^-p$, and $K^+\Sigma(1385)\pi^0$, which were simulated using an event generator with phase space distributions and a GEANT4 simulation of BGOOD, accurately implementing energy, time and spatial resolutions.  The analysis procedure of this simulated data was the same as for the identification of the $K^+\Lambda(1405)$ candidate events in the real data.

\begin{figure}[h]
	\centering
	\includegraphics[width=\columnwidth]{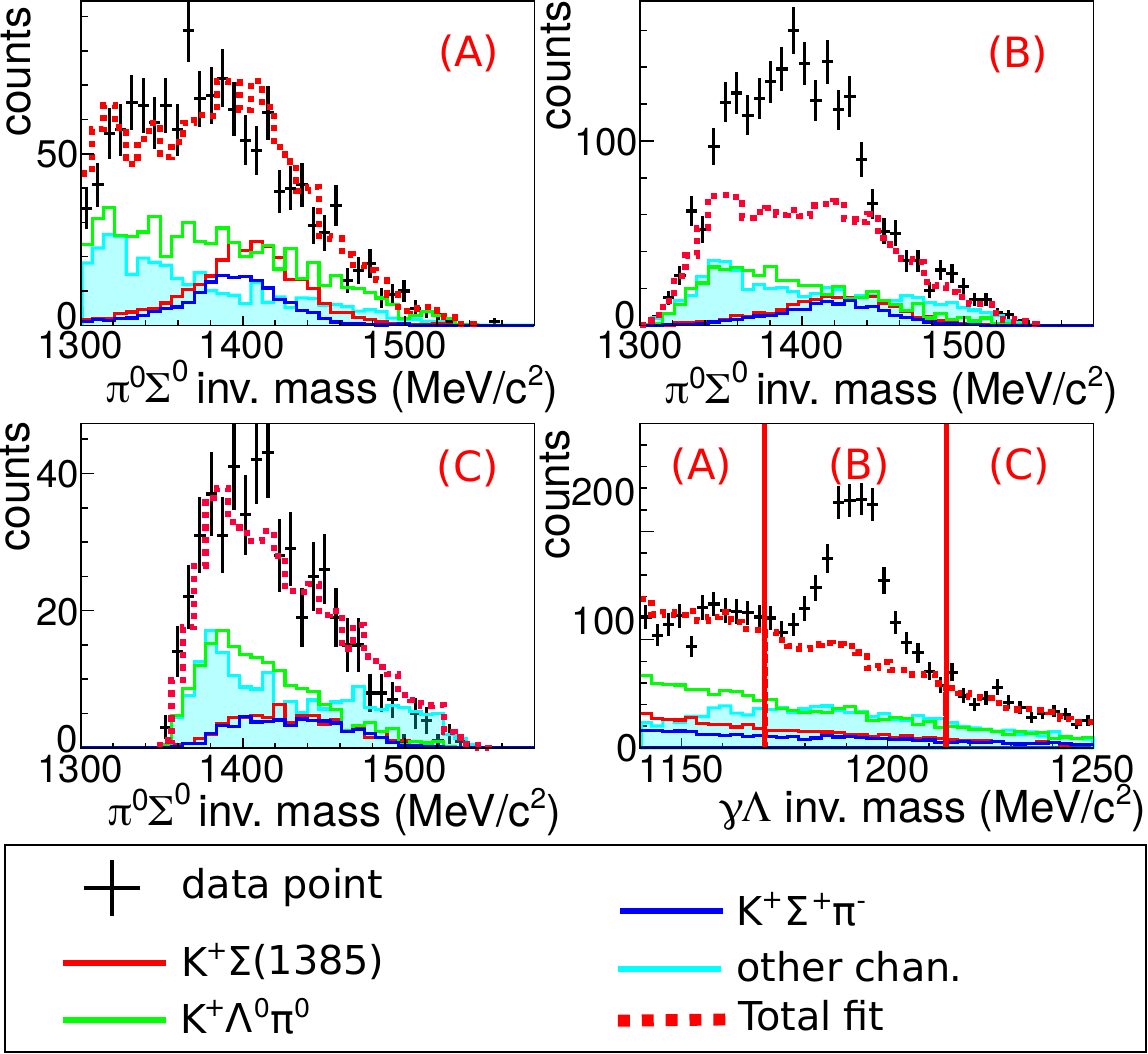} 
	\caption{One dimensional $\pi^0\Sigma^0$ invariant mass projections of the histogram in fig.~\ref{fig:RooFits_lineshape-1} for the $\gamma \Lambda$ invariant mass intervals labelled (A), (B) and (C). The lower right plot shows the $\gamma\Lambda$ invariant mass projection over the complete range.  The fit is described in the text.}
	\label{fig:RooFits_lineshape-2}
\end{figure}

The $\gamma\Lambda$ invariant mass is split into three regions labeled (A), (B) and (C) in Fig.~\ref{fig:RooFits_lineshape-1} , where the $K^+\Lambda(1405)$ signal is in the central region, (B).
The projections of the $\pi^0\Sigma^0$ invariant mass for these regions are shown in Fig.~\ref{fig:RooFits_lineshape-2}, with an additional projection of the $\gamma\Lambda$ invariant mass over all three regions.  The simulated background distributions were fitted to regions (A) and (C) where no signal was expected.
The dominant background proved to be $K^+\pi^-p\pi^0$ from the decays of $K^+\Sigma^+\pi^-$, $K^+\Lambda\pi^0$ and $K^+\Sigma^0(1385)$ final states. 
The fitted background yield from regions (A) and (C) was used to determine the background contribution in region (B) under the signal and subsequently subtracted from the data.   The remaining events were normalized according to the beam flux, detection efficiency, target area density and solid angle.

%

The differential cross section with respect to the line shape includes $K^+\Lambda(1405)$, $K^+\Lambda(1520)$ and uncorrelated $K^+\Sigma^0\pi^0$ final states.
To determine the differential cross section with respect to $E_\gamma$, the contributions of these three channels were separated.  This was achieved using two-dimensional fits to the $\pi^0\Sigma^0$ and $\gamma \Lambda$ invariant mass distributions.






The detection efficiency and resolution of the $\pi^0\Sigma^0$ invariant mass were determined using the GEANT4 model of the detector setup. 
The efficiency is almost independent of the invariant mass for masses higher than \SI{1350}{MeV/c^2}.  A loss of efficiency was determined for $\cos(\theta)$ between 0.27 to 0.42 and 0.87 to 0.89 due to small gaps between the BGO Rugby Ball, SciRi, and the Forward Spectrometer, and these data were subsequently removed from the presented results.

For  $\cos(\theta)<0.86$ the invariant mass resolution of the  $\pi^0\Sigma^0$ system varied linearly with mass and was determined as $\sigma=$\SI{15}{MeV} and \SI{17}{MeV} at  \SI{1350}{MeV/c^2} and \SI{1450}{MeV/c^2} respectively. 
For $\cos(\theta)>0.86$, the $K^+$  detection in the Forward Spectrometer improved the mass resolution
to  \SI{13}{MeV}/c$^2$ over the full invariant mass range.


Systematic uncertainties (the photon flux (4\,\%), beam energy calibration (4\,\%), hardware triggers (4\,\%), kinematic fit  (5\,\%), and the modelling of the experimental setup) were estimated from well established cross section measurements such as $\gamma p \rightarrow \pi^0/\eta p$. 
The systematic uncertainty of the detection efficiency depended on  particle angle and energy and varies between \SIrange{5}{20}{\%}. 
Differences between the simulated and real $\Lambda(1405)$ line shape for the cross section determination are included in the uncertainty of modeling the detector resolution.  The mean value for the combined systematic error is estimated as \SI{12}{\%}.


\section{Results and interpretations}

\begin{figure} [h]
	\centering
	\vspace*{0cm}
	\resizebox{1.0\columnwidth}{!}{%
		\includegraphics{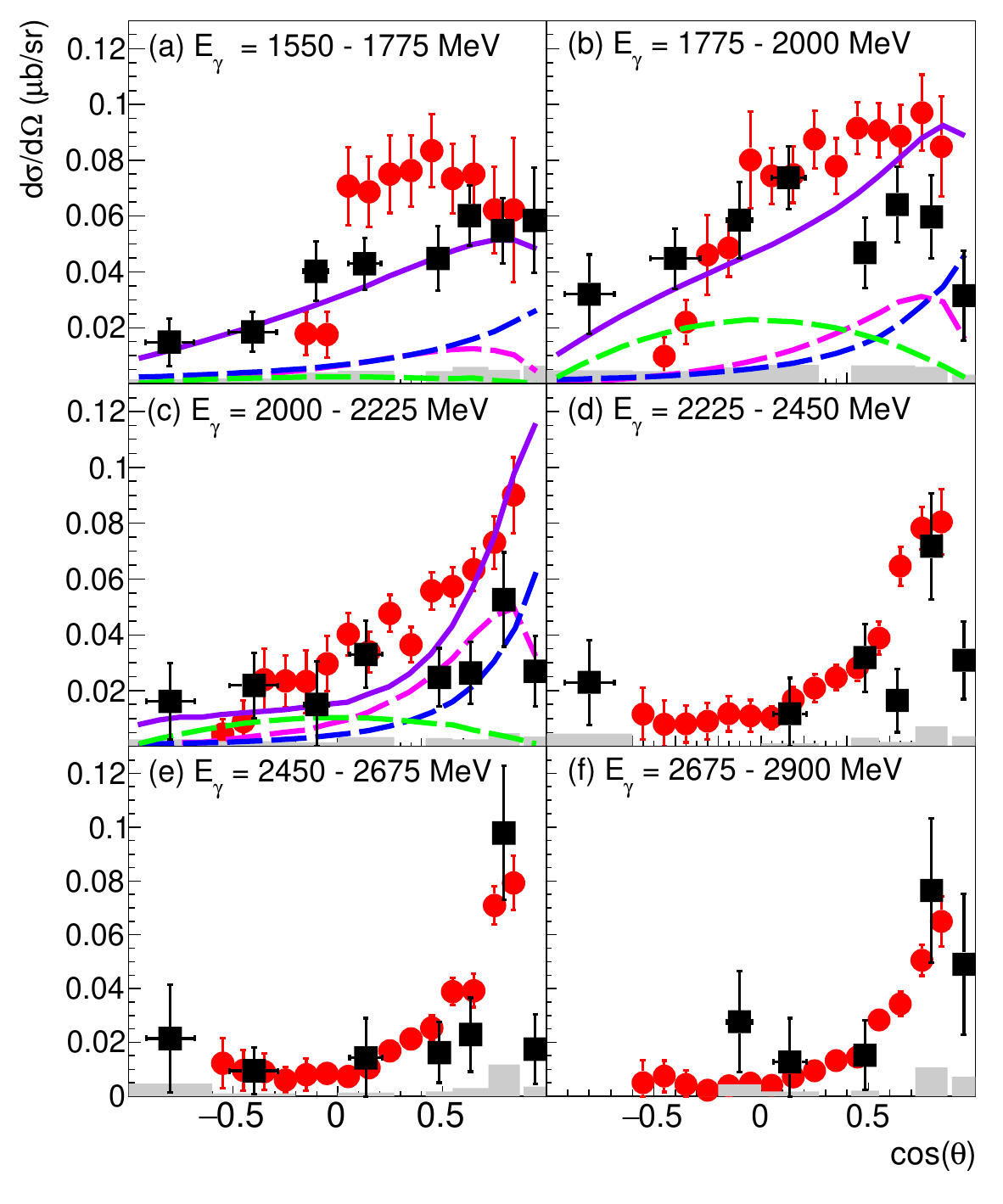} 
	}
	\caption{The $\gamma p \rightarrow K^+\Lambda(1405)\rightarrow K^+(\pi^0\Sigma^0)$ differential cross section.  The BGOOD data are the black squares with the systematic uncertainties shown as grey bars on the abscissa.  The horizontal error bars indicate  1$\sigma$ of the $\cos(\theta)$ interval.  The CLAS data~\cite{bib:clas_crosssection} are the red circles and the results of the model of Wang \textit{et al.}~\cite{bib:wang} are the superimposed lines. The dashed green line shows the triangle singularity contribution,  dashed blue line is the t-channel production via $K^*$ exchange, the dashed magenta line is $t$-channel production via $K$ exchange and the purple line is the total.}
	\label{fig:L1405_cross}
\end{figure}

The differential cross section for $\gamma p \rightarrow K^+\Lambda(1405)$ is shown in Fig.~\ref{fig:L1405_cross}.  The  data has reasonable agreement with the CLAS results and extend to previously unmeasured forward angles.
A consistency check was made by comparing $\gamma p \rightarrow K^+\Lambda(1520)$ decaying to the same final state to previous data, which also had good agreement (not shown).

The model of Wang \textit{et al}.~\cite{bib:wang} which was
fitted to the CLAS data is also shown in Fig.~\ref{fig:L1405_cross}.  The model included  t-channel production via $K$ and $K^*$ exchange, and the proposed  decay of the $N^*(2030)$ to a triangle singularity as a feeding mechanism for the $\Lambda(1405)$.  The new BGOOD data exhibits a sharp drop in the $\gamma p \rightarrow K^+\Lambda(1405)$ cross section for extremely forward $\cos(\theta)$.  The angular distribution of t-channel production via $K$ and $K^*$ are similar below $\cos(\theta)\approx0.8$, above which they start to deviate from each other. The CLAS experiment has limited coverage of forward angles, and therefore imposed no constraints on this ratio between $K$ and $K^*$ exchange.   It appears that if this new data were included in the fit, there would be an increased $K$ and smaller $K^*$ t-channel contribution, with minimal changes to the proposed triangle production amplitude.

\begin{figure} [h]
	\centering
	\includegraphics[width=\columnwidth,trim={0cm 0cm 0cm 0cm},clip]{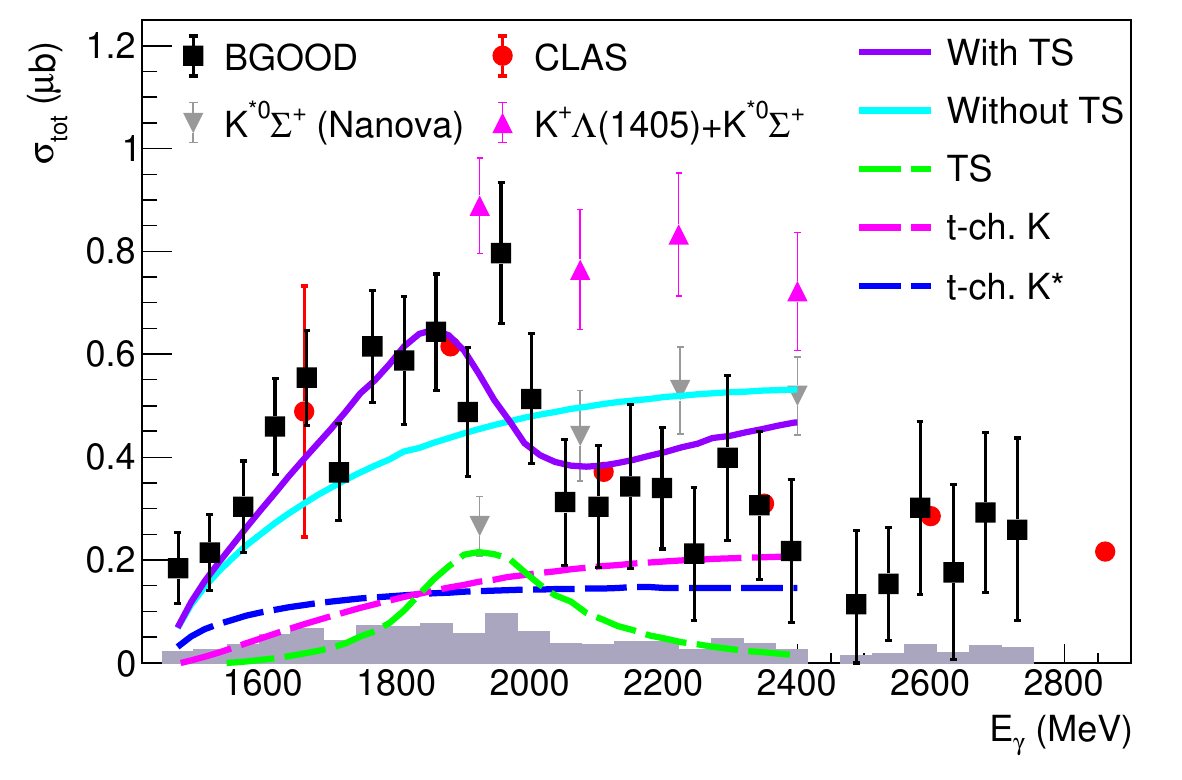} 
	\caption{Integrated $\gamma p \rightarrow K^+\Lambda(1405)$ cross section, with the same labelling as in fig.~\ref{fig:L1405_cross}.
		Additionally, the cyan line is the model of Wang \textit{et al.}~without the triangle singularity, the $K^{*0}\Sigma^+$ data from CBELSA/TAPS~\cite{bib:nanova} are the grey triangles and the sum of the $K^{*0}\Sigma^+$ and the BGOOD $K^+\Lambda(1405)$ data are the magenta triangles.}
	\label{fig:cross_total}
\end{figure}

The BGOOD detector acceptance for $K^+$ in the lab frame polar angle of \SIrange{1}{155}{\degree} translates to $\cos(\theta)\approx - 0.99$ to 0.99 for most beam energies. 
The uncovered region is therefore sufficiently small to permit the determination of the integrated cross section without extrapolation.  To do this, the number of events were weighted by the reconstruction efficiency to account for the angular dependence, which allowed the determination of the cross section with only one fit per $E_\gamma$ interval.
The integrated cross section for $\gamma p \rightarrow K^+\Lambda(1405)$ is shown in Fig.~\ref{fig:cross_total}, with the model of Wang \textit{et al.} superimposed.  The model is in agreement with both the CLAS and this new data, supporting the triangle diagram decay of  the $N^*(2030)$ resonance playing a significant role in $K^+\Lambda(1405)$ photoproduction.  The fine energy binning of this new BGOOD data reveals the extent of the cusp-like structure driven by this triangle singularity.  The falloff at $E_\gamma=2000$\,MeV is consistent with a $K^*\Sigma$ structure of the $N^*(2030)$, since in this case the triangle mechanism driven by the $N^*(2030)$ is expected to vanish once the threshold of free $K^* \Sigma$ production is exceeded.
To demonstrate this, cross section data for $\gamma p \rightarrow K^{*0}\Sigma^+$~\cite{bib:nanova} is also shown on Fig.~\ref{fig:cross_total} as the grey triangles, where it rises from threshold at the same energy of the drop in strength of the $K^+\Lambda(1405)$ cross section.  Summing the two cross sections together (the magenta triangles) gives a smooth distribution over the $K^{*0}\Sigma^+$ threshold.
This is very similar to the cusp-like structure observed in $K^0\Sigma^+$ photoproduction at the $K^*$ threshold~\cite{bib:ewald} which supports the $N^*(2030)$ as a  vector meson-baryon dynamically generated resonance in the hidden-strange sector.

\begin{figure} [h]
	\centering
	\includegraphics[width=\columnwidth,trim={0cm 0cm 0cm 0cm},clip]{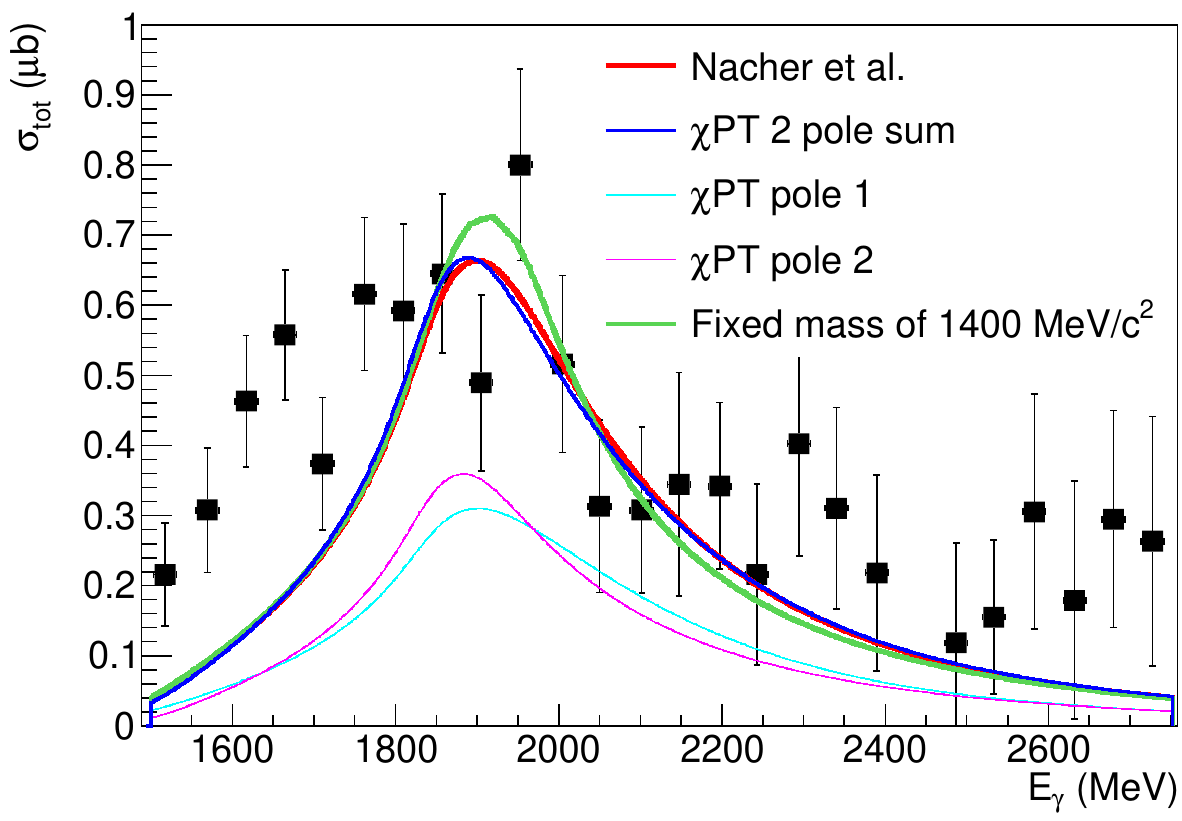}  
	\caption{Integrated $\gamma p \rightarrow K^+\Lambda(1405)$ (the same as in fig.~\ref{fig:cross_total} but with a different scale and only statistical uncertainties shown).  Superimposed are the triangle singularity calculations from the COMPASS Collaboration~\cite{mathiasthesis,alekseev10} described in the text and set at an arbitrary scale.  Indicated in the legend, three different $\Lambda(1405)$ mass distributions are used in the triangle singularity calculation:  The model of Nacher \textit{et al.}~\cite{bib:nacher} (red line), the proposed two $I=0$ poles of the $\Lambda(1405)$~\cite{bib:oller,bib:twopeak,bib:twopeakwidth} (thin cyan and magenta lines at equal amplitude and the sum as the blue line), and assuming a $\Lambda(1405)$ mass of 1400\,MeV/c$^{2}$ with no width.}
	\label{fig:cross_total_TS}
\end{figure}

Previous calculations by the COMPASS Collaboration~\cite{mathiasthesis,alekseev10}, which described the $a_1(1420)$ observed in the $f_0(980)\pi$ final state via a $\bar{K}K^*K$ triangle singularity, further support $K^+\Lambda(1405)$ photoproduction as also being driven by a triangle singularity.  Fig.~\ref{fig:cross_total_TS} shows this new data and a calculation based on the Mathematica code of Wagner~\cite{mathiasthesis} which was used to calculate the amplitude assuming the intermediate particles shown in fig.~\ref{fig:tprocess}(right).  This was multiplied by the centre-of-mass momentum squared to account for available phase space as a function of $E_\gamma$, and two additional form factors for the $N^*(2030)\rightarrow K^*\Sigma$ and $K^*\rightarrow K \pi$ vertices given in ref.~\cite{bib:wang}.  Three different $\Lambda$(1405) line shape distributions were used in the calculation and are shown in fig.~\ref{fig:cross_total_TS}.  The green line is the distribution assuming a $\Lambda(1405)$ with zero width and a mass of 140\,MeV/c$^2$, the red line uses the line shape from the model by Nacher \textit{et al.}~\cite{bib:nacher}, and the blue line is the two $I = 0$ poles predicted in $\chi$PT based models~\cite{bib:twopeak} (with equal amplitudes and also shown separately).
The distributions are at an arbitrary scale however reproduce the trend of the data well, with only small changes between the three different $\Lambda$(1405) line shapes.  Crucially, no fits to the data were made as distributions are derived exclusively from the kinematics of the triangle singularity.

\label{sec:lineshape}

\begin{figure} [H]
	\centering
	\vspace*{0cm}
	\resizebox{1.0\columnwidth}{!}{%
		\includegraphics{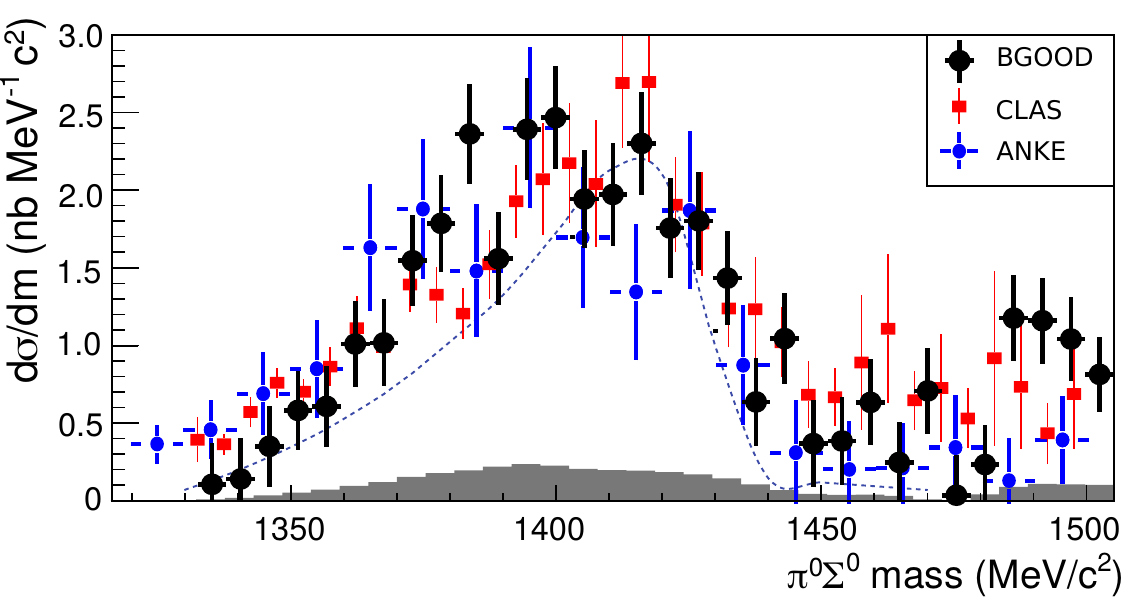} 
	}
	\caption{The $\pi^0 \Sigma^0$ invariant mass (line shape) for $E_\gamma=$ 1550 to 2300\,MeV (black circles) compared to ANKE (scaled to other data)~\cite{bib:anke_lineshape} and CLAS~\cite{bib:clas_lineshape} results (blue circles and red squares respectively). The blue dotted line is the model of Nacher \textit{et al.}~\cite{bib:nacher}. The systematic errors are the grey bars on the abscissa.}
	\label{fig:LineShapeTotal}
\end{figure}

The $\pi^0\Sigma^0$ invariant mass distribution for the full $\cos(\theta)$ range is shown in Fig.~\ref{fig:LineShapeTotal}.  The results agree with data from the CLAS collaboration~\cite{bib:clas_lineshape} within statistical errors.  The ANKE data~\cite{,bib:anke_lineshape} was not normalised but is shown here scaled to other data. 
It is interesting to note that this data and the ANKE data might indicate a two peak structure at \SI{1395}{MeV} and \SI{1425}{MeV}, which is close to the proposed two $I=0$ poles of the $\Lambda(1405)$~\cite{bib:oller}, calculated to have widths of 
\SI{132}{MeV} and  \SI{32}{MeV} respectively~\cite{bib:twopeak,bib:twopeakwidth} and in agreement with the $\chi$PT analysis based upon the CLAS $\pi^\pm \Sigma^\mp$ charged decay mode line shape measurements~\cite{bib:mai}.  These peaks however are not apparent in the CLAS data $\pi^0\Sigma^0$ line shape. 

\begin{figure} [h]
	\centering
	\includegraphics[width=\columnwidth,trim={0cm 0cm 1.5cm 0cm},clip]{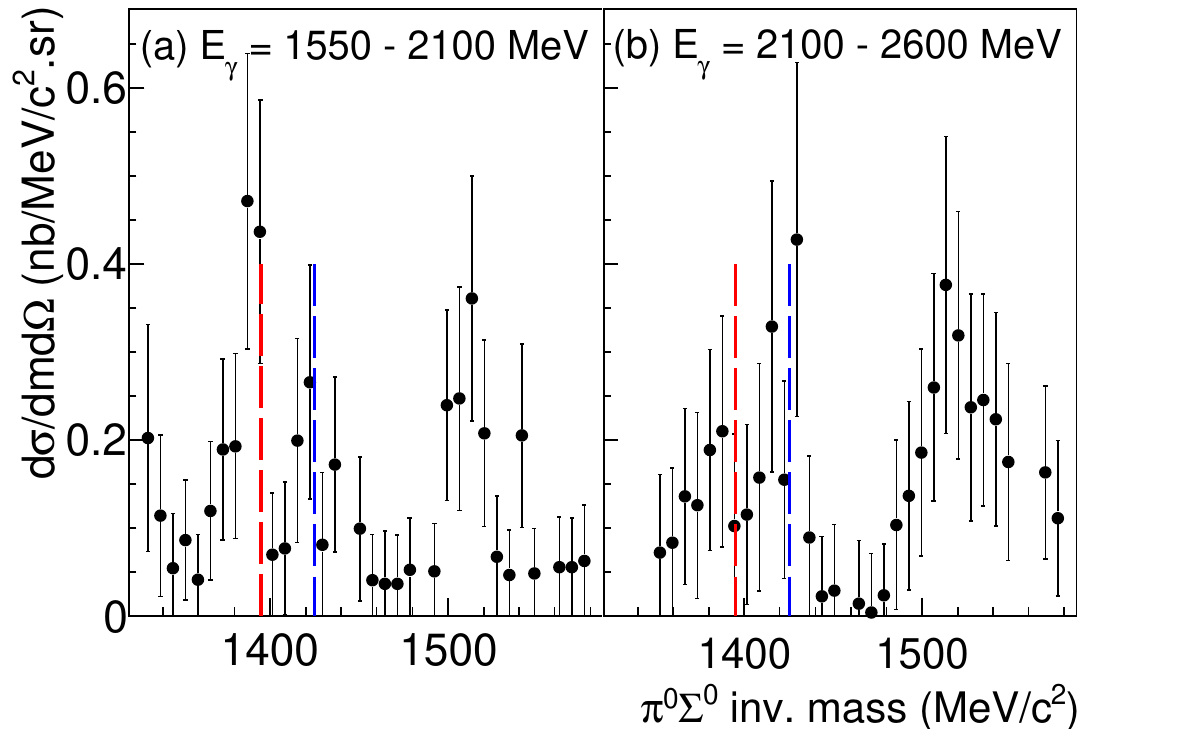} 
	\caption{The $\pi^0 \Sigma^0$ line shape  for $\cos(\theta)=$\SIrange{0.86}{1.00}{}. The dashed red and blue lines indicate the proposed poles at 1395 and \SI{1425}{MeV} respectively~\cite{bib:oller}.  }
	\label{fig:LineShapeForward}
\end{figure}

Revealed by the $K^+$ forward acceptance of this new data, these two peaks around \SI{1395}{MeV} and \SI{1425}{MeV} appear to have different cross sections at forward $\cos(\theta)$, as seen in Fig.~\ref{fig:LineShapeForward}. 
The CLAS Collaboration reported a mass resolution of \SI{7}{MeV/c}$^2$ and the two peaks were not observed~\cite{bib:clas_lineshape}, therefore, with the mass resolution of this new data of \SI{13}{MeV/c}$^2$ these peaks should also not be resolved.  CLAS had a more limited forward acceptance however, 
so it may be that there is an angular dependence on the amplitudes of the poles and that they are only apparent at forward angles.
It cannot yet be confirmed due to the statistical precision that the two poles are present in the line shape and further data is essential to resolve this structure.  It is interesting however to compare to the $\Lambda(1405)$ electroproduction data from the CLAS collaboration~\cite{fig:L1405viaelectron}.  Two peaks were observed in the $\Lambda(1405)$ lineshape at $1423 \pm 2$ and $1368 \pm 4$\,MeV/c$^2$, the relative amplitudes of which depended on $Q^2$.  If the coupling to the two poles is kinematically dependent (as is expected from $\chi$PT models), there may be an equivalence observed by variations in $Q^2$ in electroproduction and variations in the momentum exchange, $t$ in photoproduction.  Access to forward angles, and therefore low $t$ may prove to be mandatory to study such a system.

\section{Summary and conclusions}\label{sec:summary}

Results in $K^+ \Lambda(1405)\rightarrow K^+(\pi^0\Sigma^0)$ photoproduction are presented by utilising the two main features of the BGOOD experiment; the hermetic coverage to observe hyperon  decays, combined with extremely forward and high momentum resolution $K^+$ detection. 

The measured cross sections support the role of a triangle singularity driven by the $N^*(2030)$, where
$\chi$PT models of vector meson-baryon interactions determine this to be a dynamically generated state~\cite{bib:wang}.
Equivalent models successfully predicted the pentaquarks~\cite{bib:juunwu} observed in the hidden-charm sector~\cite{bib:LHCbpenta}, and this data supports the description of the $N^*(2030)$ as the hidden-strange analogue of these states~\cite{bib:OsetRamos2,bib:ramosprivate}.

The measured $\Lambda(1405)\rightarrow \pi^0\Sigma^0$ line shape is in agreement with previous measurements from the CLAS and ANKE collaborations~\cite{bib:clas_lineshape,bib:anke_lineshape}.  An indication is found for a possible double peak structure in the neutral ($\pi^0 \Sigma^0$) decay channel, consistent with the two poles derived in $\chi$PT~\cite{bib:mai}, however more data is needed for a firm conclusion.

\section*{Acknowledgements}

We thank the staff and shift-students of the ELSA accelerator for providing an excellent beam.   We thank Mathias Wagner for valuable discussion and use of his Mathematica code to calculate the triangle singularity, Volker Metag for further discussions and Eulogio Oset, \`{A}ngels Ramos and Reinhard Schumacher for insightful comments.

This work is supported by the Deutsche Forschungsgemeinschaft Project Numbers 388979758 and 405882627, the RSF grant number 19-42-04132, the Third Scientific Committee of the INFN.   This publication is part of a project that has received funding from the European Union’s Horizon 2020 research and innovation programme under grant agreement STRONG–2020 No.~824093. P.~L.~Cole gratefully acknowledges the support from both the U.S. National Science Foundation (NSF PHY-1307340, NSF-PHY-1615146, and NFS-PHY-2012826) and the Fulbright U.S. Scholar Program (2014/2015).


\appendix

\end{document}